\documentclass[twocolumn,nofootinbib,showpacs,amsmath,amssymb]{revtex4}
\usepackage{amsfonts,amssymb,amscd,amsmath}
\usepackage{graphicx}
\usepackage{subfigure}
\usepackage{bm}
\graphicspath{{figures/}}

\def\gsim{\mathrel{\rlap{\lower4pt\hbox{\hskip1pt$\sim$}}
 \raise1pt\hbox{$>$}}}

 \newcommand\la{\langle}
 \newcommand\ra{\rangle}
 \newcommand\beq{\begin{equation}}
 \newcommand\noi{\noindent}
 \newcommand\eeq{\end{equation}}
 \newcommand\beqn{\begin{eqnarray}}
 \newcommand\eeqn{\end{eqnarray}}

\def\fm{\,\mbox{fm}}
\def\GeV{\,\mbox{GeV}}
\def\TeV{\,\mbox{TeV}}
\def\lsim{\mathrel{\rlap{\lower4pt\hbox{\hskip1pt$\sim$}}
    \raise1pt\hbox{$<$}}}         
\def\gsim{\mathrel{\rlap{\lower4pt\hbox{\hskip1pt$\sim$}}
    \raise1pt\hbox{$>$}}}         

\def\fm{\,\mbox{fm}}
\def\GeV{\,\mbox{GeV}}
\def\MeV{\,\mbox{MeV}}

\begin{document}
\title{Survival of heavy flavored mesons
       in a hot medium}
\author{B. Z. Kopeliovich}
\email{boris.kopeliovich@usm.cl}
\affiliation{
Departamento de F\'{\i}sica,
Universidad T\'ecnica Federico Santa Mar\'{\i}a,
Avenida Espa\~na 1680, Valpara\'iso, Chile
}

\author{Jan Nemchik}
\email{nemcik@saske.sk}
\affiliation{
Czech Technical University in Prague, FNSPE, B\v rehov\'a 7, 11519
Prague, Czech Republic}
\affiliation{
Institute of Experimental Physics SAS, Watsonova 47, 04001 Ko\v
sice, Slovakia
}

\author{I. K. Potashnikova}
\email{irina.potashnikova@usm.cl}
\affiliation{
Departamento de F\'{\i}sica,
Universidad T\'ecnica Federico Santa Mar\'{\i}a,
Avenida Espa\~na 1680, Valpara\'iso, Chile
}

\author{Ivan Schmidt}
\email{ivan.schmidt@usm.cl}
\affiliation{
Departamento de F\'{\i}sica,
Universidad T\'ecnica Federico Santa Mar\'{\i}a,
Avenida Espa\~na 1680, Valpara\'iso, Chile
}

\begin{abstract}
Hadronization of heavy quarks reveals various unusual features.
Gluon radiation by a heavy quark originated from a hard process, 
 ceases shortly on a distance of the order of few fm.
Due to the dead-cone effect a heavy quark radiates only a small fraction of its energy. This is why the measured fragmentation function $D(z)$ peaks at large $z$.
Hadronization finishes at very short distances, well shorter than 1 fm, 
by production of a colorless small-size $Q\bar q$ dipole. This ensures
dominance of a perturbative mechanism and makes possible factorization of 
short and long distances. The latter corresponds to final state interactions of
the produced dipole propagating through a dense medium. The results provide good description of data on beauty and charm suppression in heavy ion collisions, fixing the transport coefficient for $b$-quarks about twice smaller than for charm, and both significantly lower that the values determined from
data on suppression of high-$p_T$ light hadrons. We relate this to reduction of the QCD coupling at higher scales, and suppression of radiation by the dead-cone effect.

\end{abstract}

\maketitle

\section{Introduction: ISI vs FSI}
\label{intro}
 Initial state interactions (ISI) in heavy ion collisions are usually associated with the early stage, excluding interactions with the products of the collision (partons and hadrons),
 which are attributed to final state interactions (FSI). 
 The ISI stage is subject to shadowing effects, which might essentially suppress heavy quark production. 
 The magnitude of these effects was evaluated in \cite{kt-hf}. 
 The main part of the suppression was found to come from leading twist gluon shadowing.
 The higher twist contribution, suppressed by the inverse quark mass squared, is much smaller, but still important for charm production, while is vanishingly small for production of beauty. 
 The total shadowing effect was found in \cite{kt-hf} rather strong, reducing the $p_T$-integrated charm and 
 beauty production rate at c.m. energy $\sqrt{s}=5.5\TeV$ to about a half of its original value.

 Such a strong ISI suppression of heavy flavors comes mostly from small transverse momenta of the $Q\bar Q$ pair, below the saturation scale, while the shadowing effects fade away at higher transverse momenta $p_T^2 > Q_s^2$, where the saturation scale $Q_s^2$ was evaluated in \cite{broad}.
 
 Usually, the ISI and FSI stages are assumed to factorize,
 i.e. to contribute to the cross section as a product of probabilities of two processes. 
 Frequently such an approximation fails, in particular, when the ISI stage is lasting longer 
 than the onset of the FSI stage. 
 Then the effects of coherence create links between the two stages. 
 In particular, we will show that at large transfer momenta the produced heavy quark 
 $Q$ neutralizes its color picking up a light antiquark $\bar q$, producing a colorless 
 heavy-light $Q\bar q$ dipole, at a very short time interval \cite{Lp}.
 On the other hand, the formation time of the meson wave function 
 might be long, considerably exceeding the dimension of the created dense medium. 

 Production of open heavy flavors with high transverse momenta can serve as a probe for the properties
 of the hot medium created in heavy ion collisions. The popular energy loss scenario is based on the unjustified assumption of a long hadronization length, exceeding the dimension of the medium. 
 Since radiative energy loss by heavy quarks was found significantly reduced 
 in comparison with light quarks by the dead cone effect \cite{dk}, production of heavy flavored mesons 
 was expected to be considerably less suppressed. 
 However later measurements (see e.g. \cite{b-atlas,D-alice}) did not confirm this 
 expectation, what was especially puzzling for $B$-mesons.

 We propose a natural mechanism, explaining the observed suppression of high-$p_T$
 {\bf heavy flavored mesons}, basing on study of the space-time development of the production process. 
 First of all, we conclude that the production time, i.e. the time interval starting from the hard collision producing the heavy quark, up to its color neutralization, 
 which stops energy dissipation for gluon radiation,
 \footnote{One should discriminate between vacuum and 
 medium-induced energy loss. The former is caused by regeneration of the color field of the quark, which was 
 stripped off in a hard process, while the latter is induced by broadening in the medium \cite{bdmps}.}
 is extremely short, as was demonstrated recently in \cite{Lp}.

 In the case of hadronizing light quarks, the time of color neutralization 
 is quite short \cite{jet-lag} and the produced colorless $q\bar q$ dipole has to survive propagating through the dense medium before being projected to the wave function of the final hadron.
 Thus, color transparency, rather than induced energy loss, becomes the major phenomenon controlling the FSI 
 effects in production of high-$p_T$ light hadrons in $AA$ collisions
 \cite{within,jet-lag,ct,ct-eloss}.

 However, for heavy flavor production the suppression mechanism undergoes
 essential modifications. Survival of the produced $Q\bar q$ is not an issue anymore, 
 because after a break-up the heavy quark $Q$ easily picks-up another light $\bar q$ 
 without any essential loss of energy, carried mainly by the heavy quark. 
 Thus, the survival probability of a $Q\bar q$ dipole is close to unity, and it is easily 
 transpassing the dense medium up to the periphery. 

 This paper is organised as follows.
Section \ref{eloss} presents the main features of heavy quark hadronization in vacuum. The dead-cone effect leads to a rather short time of radiation of the full spectrum of gluon, which take away only a small fraction of the quark energy. 
In section \ref{L_p} we relate the production length distribution directly to the measured fragmentation function. The production length turns out to be exceptionally short, much shorter that $1\fm$. This observation allows to factorize in-medium hadronization to short-length perturbative stage and long-range final state interaction in the medium. The latter is studied in section \ref{medium}. The crucial features of a heavy-light dipole propagating through a medium, which affect the resuls, are fast expansion and large size of heavy flavored mesons. The results are compared with data in section \ref{data}, allowing to determine the transport coefficient, which turnes out to be much smaller for $B$- than for $D$-production. And both are considerably smaller than has been found for light hadron. This peculiar result is interpreted in section \ref{q-hat}.

\section{Hadronization of heavy quarks in vacuum}
\label{eloss}
\subsection{Regeneration of the color field}
 A parton originated from a hard reaction is lacking a color field with transverse momenta (relative to the quark momentum) up to the hard scale of the process, 
 $k_T^2<Q^2=q_T^2+m_Q^2$, 
 where $q_T$ and $m_Q$ are the  quark transverse momentum and mass respectively.
 Regeneration of the color field is accompanied by gluon radiation, forming a jet of hadrons. 
 These gluons are carrying a part of the total jet energy, and since
 the radiation process has time ordering, it can be treated as energy dissipation, or vacuum energy loss.
 The coherence length of gluon radiation by a heavy quark of energy $E$ has the form,
\beq
L^g_c=\frac{2E\,x(1-x)}{k_T^2+x^2\,m_Q^2}\,,
\label{140}
\eeq
 where $x$ is the fractional light-cone (LC) momentum of the radiated gluon, 
 and $k_T$ is its transverse momentum relative to the jet axis. 
 Thus, gluons are radiated sequentially, rather than burst simultaneously. 
 First  of all are radiated gluons with small 
 longitudinal and large transverse momenta, while gluons with small 
 radiation angles get to mass shell later.

 The peculiar feature of radiation by a heavy quark is the quick regeneration of the field, compared with light quarks.
 This is related to the specific radiation spectrum,
\beq
\frac{dn_g}{dx\,dk_T^2} =
\frac{2\alpha_s(k_T^2)}{3\pi\,x}\,
\frac{k_T^2[1+(1-x)^2]}{[k_T^2+x^2m_Q^2]^2}\,.
\label{145}
\eeq
 Remarkably, while the radiation spectrum of light quarks peaks
 (is divergent) at small $k_T$, the spectrum Eq.~(\ref{145})
 excludes gluons with small $k_T\lesssim xm_Q$. This is known in the literature as \textit{dead cone effect} 
 \cite{troyan}. Coherence length (\ref{140}) of such large-$k_T$ gluons is short, while gluons with small $k_T$,  which have long production length are suppressed. Therefore we expect short time of radiation and prompt regeneration of the field by heavy quarks. Moreover, since considerable part of the radiated energy is carried by low-$k_T$ gluons, which are suppressed, only a small fraction of the quark energy is radiated, in contrast to light quark jets. These distinctive features of heavy quarks can be checked by direct calculations.

 The amount of energy, radiated along the quark path length $L$ 
 from the hard collision point can be evaluated imposing a requirement that
 only those gluons contribute, whose radiation length Eq.~(\ref{140}) is shorter than $L$.
 It is given by the following integral of the radiation spectrum Eq.~(\ref{145}) \cite{knp,within,similar},
\beq
\Delta E_{rad}(L) =
\int\limits_{\lambda^2}^{Q^2}
dk_T^2\int\limits_0^1 dx\,\omega\,
\frac{dn_g}{dx\,dk_T^2}\,
\Theta(L-L^g_c)\,,
\label{130}
\eeq
 where $\omega$ is the gluon energy.

\begin{figure}[hbt]
\vspace{-0.2cm}
    \includegraphics[height=4.2cm]{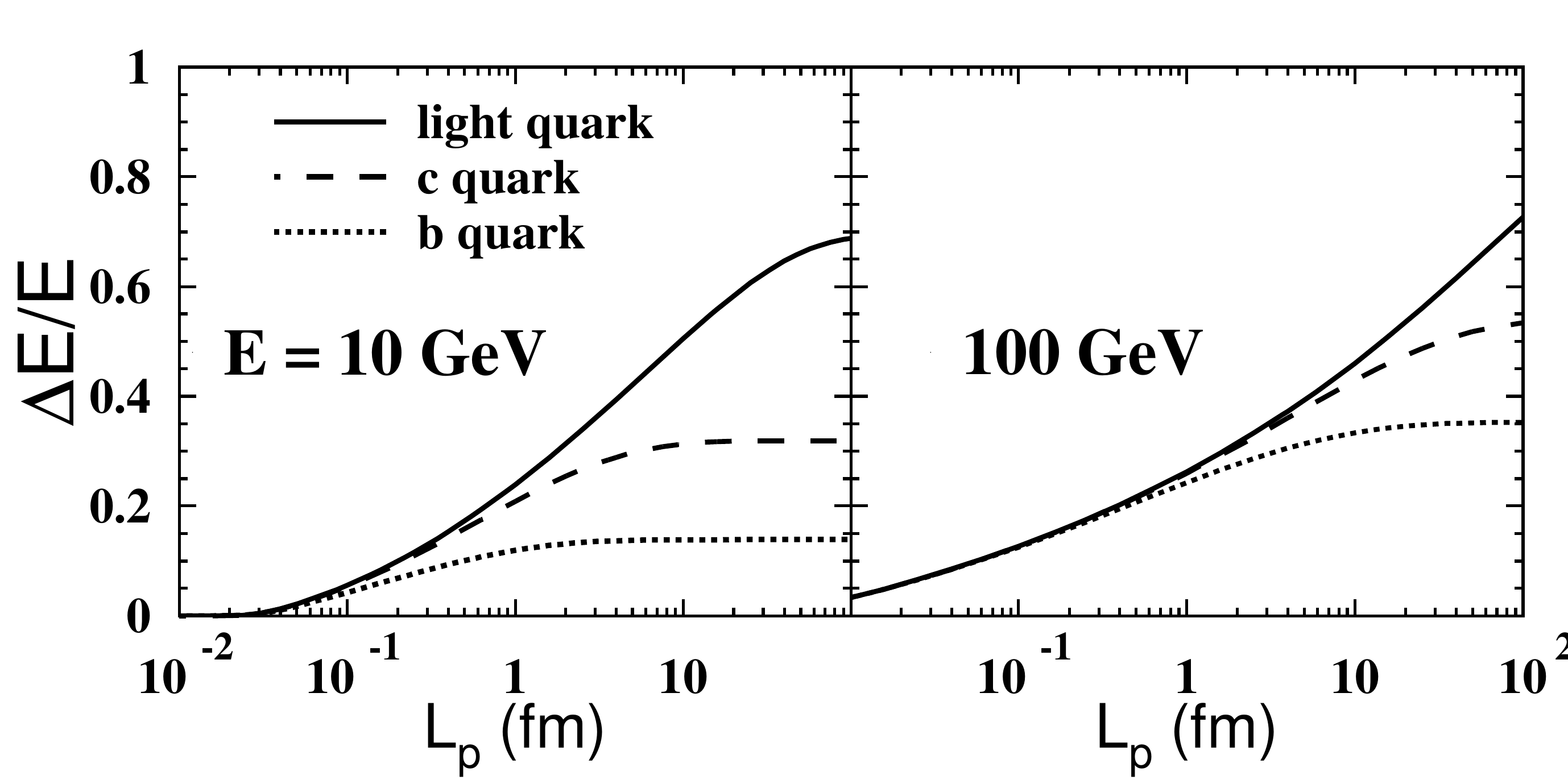}
    \caption{ \label{fig:eloss-qn}
         Fractional radiational  energy loss in vacuum by light, $c$ and $b$ quarks 
         as function of path length for  
         different initial quark energies.}
\end{figure}

 Examples of length-dependent fractional radiated energy for light and heavy  quarks are shown in Fig.~\ref{fig:eloss-qn}.
 We see some specific features of $L$-dependence of the fractional radiational energy loss in vacuum: 
 (i) in contrast to light quarks, which continue radiating long time over hundred Fermi, 
 radiation of heavy quarks ceases shortly on a distance of the order of $1\fm$ (depending on energy). 
 This means that heavy quarks quickly complete the field restoration process. 
 (ii) The full fraction of radiated energy by heavy quarks is rather 
 small. E.g. for $b$-mesons this faction ranges from 10 to 35 \% for quark energies 10-100\,GeV.

 These plots also reveal another peculiar feature of vacuum radiation.
 In spite of expectation, at the early stage of radiation process all species of quarks radiate universally. This is a consequence of the step function $\Theta(L-L^g_c)$ in Eq.~(\ref{130}), which
 creates another, even wider dead cone \cite{similar}, 
 leading to a suppression of gluon radiation with sufficiently low transverse momenta,
\beq
k_T^2 <
\frac{2 E x (1 - x)}{L} - x^2 m_Q^2\,.
\label{148}
\eeq
 This bound is practically independent of the quark mass, so heavy and light quarks radiate similarly for a while. However, the bound (\ref{148}) relaxes gradually with time, reaching the magnitude
 $k_T^2\approx x^2 m_q^2$ characterizing the heavy quark dead cone, at the distance
\beq
L = 
\frac{E (1 - x)}{x m_Q^2}\,.
\label{150}
\eeq
 At longer path length the dead cone related to the heavy quark
 mass becomes effective, and the heavy and light quarks start radiating differently.  
 The characteristic distance Eq.~(\ref{150}) for $b$ quarks is an order of magnitude
 shorter than for $c$, and both rise linearly with jet energy. 

\subsection{Production of heavy-flavored mesons}\label{L_p}

 For the sake of concreteness, in what follows we consider mostly $b$-quarks and $B$-mesons.  
 Extension to $D$ mesons is straightforward.

 As long as only a small fraction of the initial energy is radiated by a heavy
 quark, the final $B$ ($D$) mesons carry almost the whole momentum of the jet. 
 So the fragmentation function $D_{b/B}(z)$ should be concentrated at large values of $z$, 
 which is the fractional LC momentum of the $B$ meson.

\begin{figure}[b]
\centering
\includegraphics[width=8.5cm,clip]{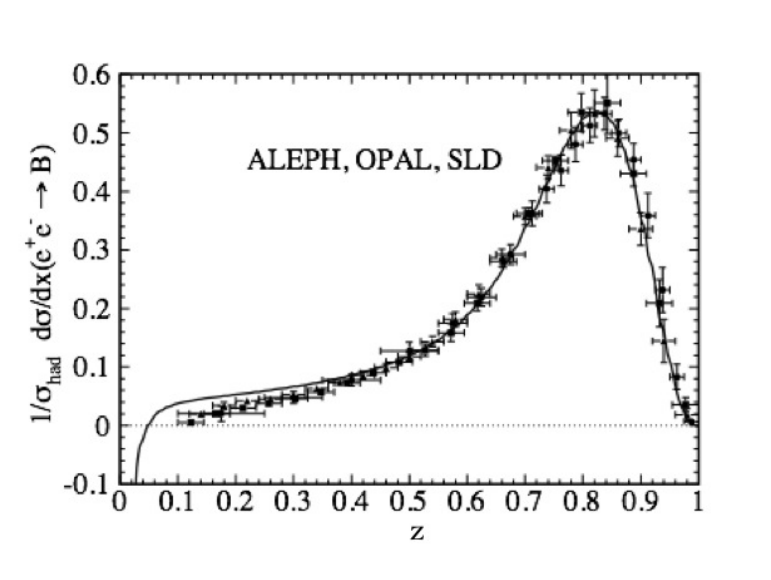}
\caption{The $b\to B$ fragmentation function, 
         from $e^+e^-$ annihilation. 
         The curve is the DGLAP fit \cite{bottom}.}
\label{fig:ff}  
\end{figure}

 This expectation is confirmed by the direct measurements of the fragmentation functions
 $c\to D$ and $b\to B$ in $e^+e^-$ annihilation \cite{charm,bottom}.
 An example of the $b\to B$ fragmentation function is depicted in Fig.~\ref{fig:ff}.
 Indeed, it shows that the distribution strongly peaks at $z\sim0.85$.
 On the contrary, the fragmentation functions of light quarks to light mesons are known 
 to fall with $z$ steadily and steeply from small $z$  to  $z=1$ \cite{kkp}.

 If a $B$ meson (in fact a small-size $b\bar q$ dipole) is produced at the distance $L_p$, 
 called production length, its momentum is preserved and the final $B$-meson acquires this momentum, 
 which is nearly equals to the momentum of the $b$-quark.
 Indeed, the fractional LC momentum of the light antiquark is very small, 
 $\alpha\approx m_q/m_b$, i.e. about $5\%$.
 The colorless dipole produced at $L=L_p$ stops radiating and does not dissipate energy any more. 

 As far as the rate of radiational vacuum energy loss $dE/dL$ is known, Eq.~(\ref{130}), one
 can relate the differential production length distribution $\partial W/\partial L_p$ 
 to the $b\to B$ fragmentation function $D_{b/B}(z)\equiv \partial W/\partial z$,
\beq
\frac{dW}{dL_p}=
\left.\frac{\partial \Delta p_+^b(L)/p_+^b}{\partial L}\right|_{L=L_p}
\!D_{b/B}(z)\, ,
\label{155}
\eeq
 where the production length distribution and the fragmentation function are normalized to unity,
 $\int_0^\infty dL_p\, dW/dL_p=1$ and  $\int_0^1 dz\,D_{b/B}(z)=1$ respectively. 
 The fractional LC momenta are defined as, 
 $z\equiv {p_+^B}/{p_+^b} = 1-{\Delta p_+^b(L_p)}/{p_+^b} = 1-\Delta z(L_p)$; and
\beq
\Delta p_+^b(L_p)=\int\limits_0^{L_p} dL\, \frac{dp_+^b(L)}{dL}\, .
\label{158}
\eeq

\begin{figure}[t]
 \vspace{-0.7cm}
\centering
\includegraphics[width=8.5cm,clip]{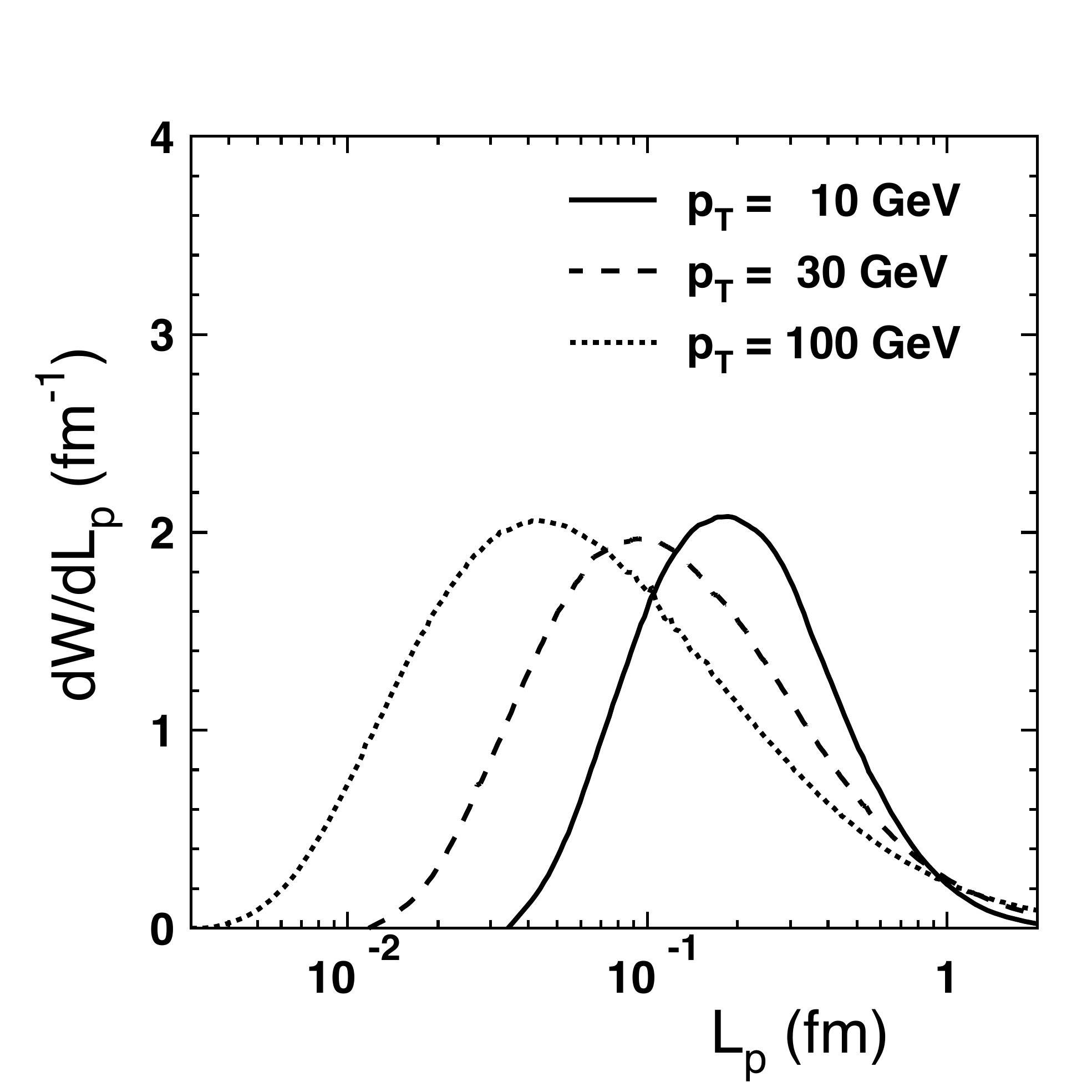}
\caption{The $L_p$-distribution of $B$-mesons produced 
         with different $p_T$ in $pp$ collisions.}
\label{fig:lp}
\end{figure}

 The rate of LC momentum loss is related to energy loss in accordance with $p_+^b=E+\sqrt{E^2-m_b^2}$.
 Thus, one can extract the production length distribution directly from data for $D_{b/B}(z)$ 
 calculating the shift of fractional momentum $\Delta z(L)$ acquired on the path length $L$. 
 The resulting production length distribution $dW/dL_p$  vs $L_p$ is 
 depicted in Fig.~\ref{fig:lp} for different transverse momenta of $B$ mesons.
 Fig.~\ref{fig:lp} 
 clearly demonstrates that
 the mean value of $L_p$ decreases with rising $p_T$,
 like it also happens for production of high-$p_T$ light hadrons \cite{jet-lag}.

\section{In-medium fragmentation}\label{medium}
 Amazing shortness of the production length $L_p$ in $pp$ collisions, demonstrated in Fig.~\ref{fig:lp} guarantees that the medium does not affect the $L_p$ distribution in heavy ion collisions. The medium
 is not created momentarily at the moment of hard $NN$ collision producing heavy quarks. 
 The characteristic time of medium creation is usually estimated at $t_0\sim 0.5-1\fm$.
 This implies factorization of short and long distances in the cross section of $B$ production 
 in collisions of nuclei (taken identical for simplicity) with impact parameter $\vec s$,
\beqn
&&
\frac{d^2\sigma_{AA\to BX}}{d^2p_Td^2s}
=
\int d^2\tau\, T_A(\tau) T_A(\vec s-\vec\tau)
\nonumber\\&\times&
\int d^2p^\prime_{T}\,
\frac{d^2\sigma_{NN\to BX}}{d^2p^\prime_{T}}\,
 V_{b\bar q\to B}(p^\prime_T,p_T,\vec s,\vec\tau,\phi)\,,
\label{215}
\eeqn
 where $T_A(s)=\int_{-\infty}^\infty dz\, \rho_A(s,z)$ is the 
 integral of the nuclear density $\rho(\vec r)$ along the collision axis. 
 The differential cross section of $B$ and $D$-meson production in $pp$ collisions was calculated in \cite{roman}.

 Factorization property explicitly divides the two stages: 
 (i) production process of a $b\bar q$ dipole, and 
 (ii) the transition probability of the $b$-quark
 $V_{b\bar q\to B}(p^\prime_T,p_T,\vec s,\vec\tau,\phi)$ through the medium. 
 The $b$-quark starts propagation as a constituent of
 the $b\bar q$ dipole, and after propagation and interaction in the medium,  
 finally ends up as the detected $B$ meson with transverse momentum $P_T$.

\subsection{Transition probability $\bf V_{Q\bar q\to B}$}

\begin{figure}[h]
\centering
\includegraphics[width=4.0cm,clip]{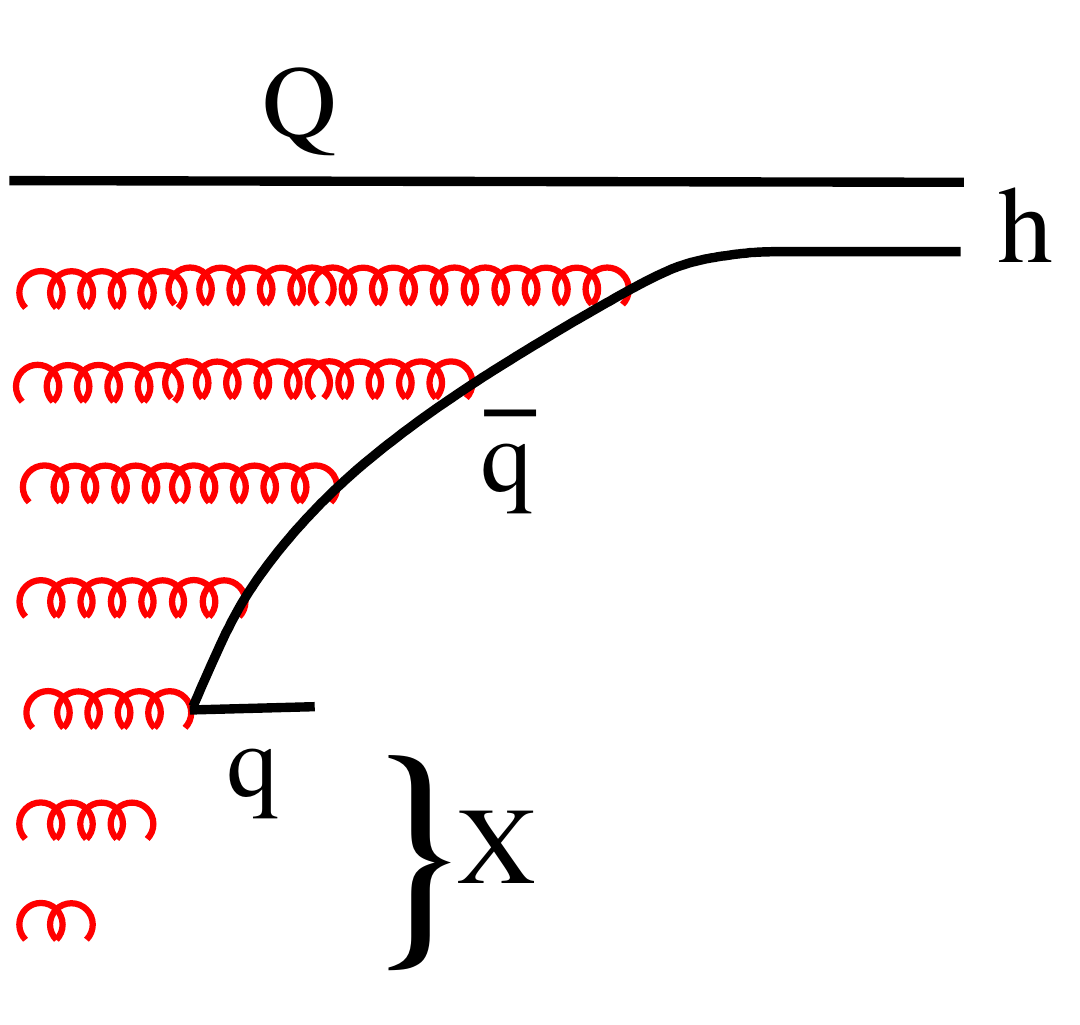} 
\vspace{-0.4cm}
\caption{Redistribution of the energy inside the $Q\bar q$
         dipole. The gluons radiated by $Q$ are absorbed by $\bar q$
         so the dipole energy remains unchanged.} 
\label{fig:regge-cut}
\end{figure}

 As was explained earlier, the $b\bar q$ dipole, crated at the distance $L=L_p$ from the production point, stops radiating gluons and propagates in vacuum without energy loss. This is why $p_T$ of the produced $B$-meson coincides with the initial $p_T$ of the $b\bar q$ dipole.
 However, at the early stage of $b\bar q$ evolution the $b$ quark is still virtual 
 and keeps radiating gluons, regenerating its color field. However, this radiation is absorbed by the co-moving antiquark $\bar q$, so that the dipole momentum remains unchanged. 
 Such a process of intrinsic radiation is illustrated in Fig.~\ref{fig:regge-cut}.

 Apparently, this process corresponds to the unitarity cut of a $q\bar q$ Reggeon in elastic scattering 
 of a $B$-meson, and the radiated gluons give rise to reggeization of the $q\bar q$ exchange. Thus, the $b$-quark, being a constituent of the $b\bar q$ dipole, keeps losing energy like in vacuum, sharing it with the accompanying light  antiquark.

 Fig.~\ref{fig:eloss-qn} shows that energy loss by a $B$-quark as function of path length ceases shortly, and the quark propagates further like a free particle without radiation. 
 Of course such a behavior contradicts confinement, which was missed within the perturbative treatment of energy loss. We add nonperturbative energy loss basing on the string model, 
 which provides the rate of energy loss $dE_{str}/dL = -\kappa$,
 where $\kappa \approx 1\GeV/\fm$ is the string tension in vacuum.
 It also participate in sharing the energy inside a $b\bar q$ dipole among the constituents.

 The full rate of energy loss due to both perturbative and nonperturbative mechanisms reads,
\beq
\frac{dE}{dL}= \frac{dE_{rad}}{dL}-\kappa\, .
\label{111}
\eeq
 However, the magnitude of the string tension, and even its very existence, depends on
 the medium temperature. We rely on the model \cite{string1,string2,string3} 
 based on the lattice simulations for temperature dependence of the string tension,
 $\kappa(T)=\kappa\left(1-T/T_c\right)^{1/3}$, where the critical temperature is fixed at $T_c=280\MeV$.
 The medium cools down with time and towards the medium periphery, so the string tension rises up to its vacuum value.

 The small-size $b\bar q$ dipole is promptly evolving to lager sizes.
 Such a dipole has no certain mass, but can be expanded over the eigenstates of the mass matrix, 
 i.e. physical states with certain masses ranging from very heavy $M\sim\sqrt{m_b^2+p_T^2}$ down to $m_B$.

 As was mentioned above, a heavy-light dipole is very asymmetric, the light quark carries a tiny fraction of the dipole momentum. This means that while the heavy quark momentum is directed forward, 
 the light slow quark is directed at a large angle, causing a rapid size expansion. 
 This is confirmed later by Eq.~(\ref{115}).  The size of a $Q\bar q$ meson, 
 controlled by the light quark mass $m_q$, is large, $r\sim 1/m_q$. E.g. the $B$-meson is nearly
 as big as pion, $\la r_{ch}^2\ra_B = 0.378\fm^2$ \cite{radius}.   

 Apparently the mean free path of such a large dipole in a dense medium is short. Namely, the mean free path is given by
\beq
L_{free} =\frac{2}{\hat q\, \la r_T^2\ra}\,,
\label{255}
\eeq
 where $ \la r_T^2\ra = 8\la r_{ch}^2\ra/3$. The so called transport coefficient $\hat q$ is the rate
 of broadening quark transverse momentum in the medium (see Sect.~\ref{h-hat}).

 As an example, for a typical value of $\hat q = 1\,\GeV^2/\fm$ 
 the mean free path of a $B$-meson is  $L_{free}=0.04\fm$, i.e. the $b$-quark
 propagating through a hot medium, frequently loses the co-moving light antiquarks, and picks up a new one, over and over again. Important is that $b$ keeps the same energy loss, whether it is free, or inside a colorless dipole. After the last breakup and recreation at distance $L^\prime_p$ 
 from the original production  point, the $b\bar q $ dipole survives,
 propagating through the rest of the medium and is developing the $B$-meson wave function. 

 Apparently, this happens mostly on the diluted medium periphery,
 therefore 
 $L^\prime_p$
 in general is rather long, several Fermies.
 The total energy loss is given by
\beq
\Delta E(L^\prime_p)=\int\limits_0^{L^\prime_p} dL\,
\frac{dE}{dL}\,,
\label{112}
\eeq
 where the rate of energy loss is given by Eq.~(\ref{111}).
 Correspondingly, the momentum loss is $\Delta p_T=\sqrt{(E_T+\Delta E)^2-m_b^2}-p_T$, where $E_T=\sqrt{p_T^2+m_b^2}$. 

 Thus, the transition of the $b\bar q$ dipole, primarily produced in an $NN$ hard collision, 
 into the detected $B$-meson, is strongly affected by propagation through the medium, 
 which causes attenuation of the dipole, as well as energy loss. Correspondingly, the transition probability can be represented as,
\beqn
V_{b\bar q\to B}(\vec p^{\,\,\prime}_T,\vec p_T,\vec s,\vec\tau,\phi)&=&
V_1(\vec p^{\,\,\prime}_T,\vec p_T,\{...\})\nonumber\\ &+&
V_2(\vec p^{\,\,\prime}_T,\vec p_T,\{...\}),
\label{V}
\eeqn
 where the term $V_1$ describes the rare situation when the primarily produced $b\bar q$ dipole manages to 
 propagate through the medium without interactions and to switch into a $B$-meson with no momentum loss.
 Correspondingly,
\beqn
&&
V_1(\vec p^{\,\,\prime}_T,\vec p_T,\vec s,\vec\tau,\phi) =
\delta^{(2)}(\vec p^{\,\,\prime}_T-\vec p_T) \nonumber\\&\times&
\exp\left[-{1\over2}\int\limits_0^\infty
dl\, r^2_{b\bar q}(l)\,\hat q(l,\{...\})\right].
\label{V1}
\eeqn
 Here $\hat q$ is the transport coefficient, the parton broadening rate,
 which depends on the path length $l$ and geometry of the collision and the trajectory, $\vec s,\vec\tau,\phi$. 
 The transverse dipole separation squared,
 $r^2_{b\bar q}(l)$ is evolving as function of path length, starting from the small initial size, $r^2_{b\bar 
 q}(L_p)\sim 1/(E_T^\prime)^2$, where $E_T^\prime=\sqrt{(p^\prime_T)^2+m_b^2}$.
 Then it rises as \cite{ct-eloss},
\beq
r^2_{b\bar q}(l)=
\frac{2l}{\alpha(1-\alpha)E_T^\prime} +
\frac{1}{(E_T^\prime)^2}.
\label{115}
\eeq
\\ 
\noi
At small fractional LC momentum of the light quark, $\alpha\approx m_q/m_Q$, the characteristic 
expansion (formation) length is,
\beq
L_f\approx \frac{m_q\,\sqrt{p_T^2+m_Q^2}}{2m_Q}\,
\la r_M^2\ra,
\label{Lf}
\eeq
 where $\la r_M^2\ra$ is the mean square radius of either $B$ or $D$-mesons.
At this distance the dipole separation reaches the meson size, and we freeze this dipole size at longer distances in Eq.~(\ref{V1}).

 The second term in (\ref{V}) corresponds to one or more inelastic interactions
 of the $b\bar q$ dipole transpassing the medium. 
 The last interaction and dipole breakup occurs at a distance $L^\prime_p$ from the origin.
 Correspondingly,
\begin{widetext}
\beqn
V_2(\vec p^{\,\,\prime}_T,\vec p_T,\vec s,\vec\tau,\phi)
={1\over2}
\int\limits_0^\infty dL^\prime_p\,
 r^2_{b\bar q}(L^\prime_p)\,
\hat q(L^\prime_p,\{...\})\,
\delta^{(2)}\left(\vec p^{\,\,\prime}_T-\vec p_T-\Delta \vec p_T\right)
\exp\left[-{1\over2}\int\limits_{L^\prime_p}^\infty
dl\,r^2_{b\bar q}(l)\,\hat q(l,\{...\})\right],
\label{V2}
\eeqn
\end{widetext}
 where $\Delta p_T$ is the momentum loss, defined above.

\section{Comparison with data}\label{data}
 Now we are in a position to calculate the suppression factor $R_{AA}(\vec{s})$
 of heavy flavored mesons produced with high $p_T$ in a nuclear collision 
 with relative impact parameter $\vec s$,
\beqn
R_{AA}(\vec{s},\vec p_T) 
=
\frac{d^2\sigma_{AA}(s)/d^2p_Td^2s}
{T_{AA}(s)\,d^2\sigma_{pp}/d^2p_T},
\label{850}
\eeqn
 where $T_{AA}(s)=\int d^2\tau\,T_A(\tau)T_A(\vec s-\vec\tau)$;
 the differential cross sections $d^2\sigma_{pp}/d^2p_T$
 and $d^2\sigma_{AA}/d^2p_T$ are given by Eq.~(\ref{215}).

 The differential cross section of $B$-meson production, $d^2\sigma_{pp}/d^2p_T$ was successfully calculate within the dipole approach in \cite{roman}. However we prefer here to rely on data and 
 fitted to ATLAS \cite{atlas-B-pp} and CMS \cite{cms-B-pp} 
 experimental values
 as
\beq
\frac{d\sigma_{pp}}{dp_T} \propto
\frac{1}{(p_T^2+m_B^2)^{\alpha_B}}
\label{880}
\eeq
 with $\alpha_B=2.7\pm 0.1$.

Another option for getting the cross section of $B$ production would be to employ available data on 
$B$-jets \cite{jet-b}, which give direct access to $q_T$ distribution of produced $b$-quarks,
$d\sigma/dq_T \propto 1/(q_T^2 + m_b^2)^{\alpha_b}$ with $\alpha_b=2.55\pm0.08$. Data for $D$-jets from \cite{jet-c1,jet-c2}, parametrized by the same way, give $\alpha_c=2.64\pm0.12$.

As far as the quark production cross section is known, it is streightforward to
calculate the meson $p_T$ distribution,
\beq
\frac{d^2\sigma_{pp\to B}}{d^2p_T}=\frac{1}{2\pi p_T E_T}
\int d^2 q_T\,\frac{d^2\sigma_{pp\to b}}{d^2q_T}\,
z\,D_{b/B}(z).
\label{890}
\eeq

 Analogously, for production of $D$-mesons in $pp$ collisions data of ALICE \cite{alice-D-pp} and CMS 
 \cite{cms-D-pp} were fitted with parametrization
 similar to (\ref{880}), $d\sigma_{pp}/dp_T\propto / (p_T^2 + m_D^2)^{\alpha_D}$ with
 $\alpha_D=2.6\pm 0.1$.

We found excellent agreement of the directly measured and calculated cross section of $B$ and $D$ production.

 For the transport coefficient, i.e. the broadening rate, along the trajectory of the produced $b$-quark through the dense medium, we employ the popular model \cite{wang},
\beq
\hat q(l,\vec s,\vec\tau,\phi)=\frac{\hat
q_0\,t_0}{t}\, \frac{n_{part}(\vec s,\vec\tau + l\,\vec p_T/p_T)}{n_{part}(0,0)}
\,\Theta(t-t_0)\, ,
\label{900}
\eeq
where 
 $n_{part}(\vec s,\vec\tau)$ is the number of participants at transverse coordinates
 $\vec s$ and $\vec \tau$ relative to the centers of the colliding nuclei.
 The falling time dependence, $1/t$ is due to longitudinal expansion of the produced medium. 
 The time interval $t_0$ required for equilibrated medium production is poorly known, 
 and even not well defined. We fixed it at the frequently used value $t_0=1\fm$.
 The path length $l$ is related to the $B$-meson speed, $l=v_Bt$, where 
 $v_B=p_T/\sqrt{p_T^2+M_B^2}$ is the $B$-meson transverse velocity.

 As far as we defined all the quantities, required for the nuclear ratio Eq.~(\ref{850}),
 the calculation becomes parameter-free, except the parameter $\hat q_0$, which is
 the maximal value of the transport coefficient at $s=\tau=0$ and $t=t_0$. In fact, measurement of this parameter is the goal of such an analysis.

 Medium-induced energy loss, including radiative and collisional mechanisms \cite{baier}, 
 are added as well, although they are considerably smaller than vacuum radiative energy loss, especially for 
 heavy quarks.

\begin{figure}[h]
\centering 
\includegraphics[width=8.0cm,clip]{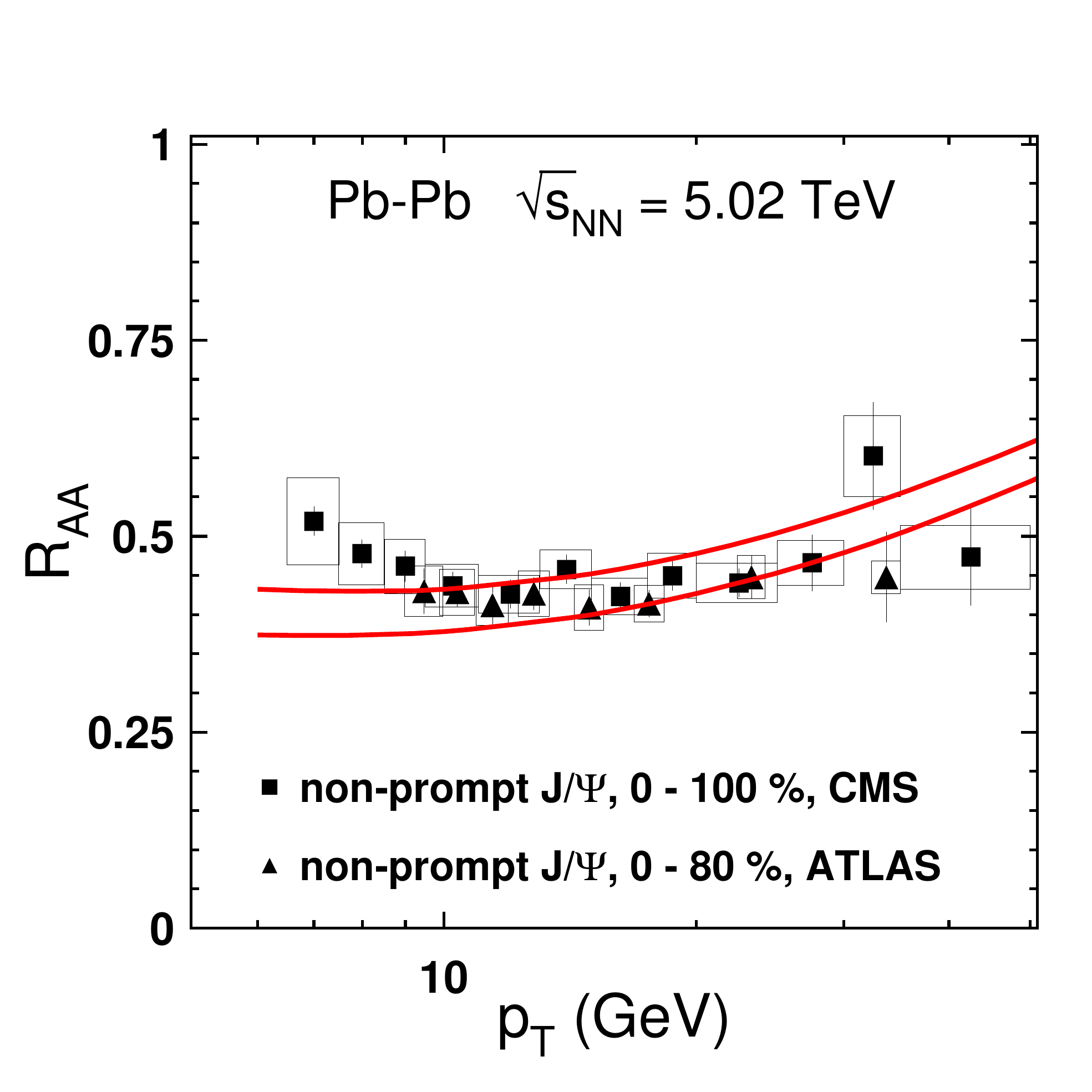}\vspace{-0.2cm}
\includegraphics[width=8.0cm,clip]{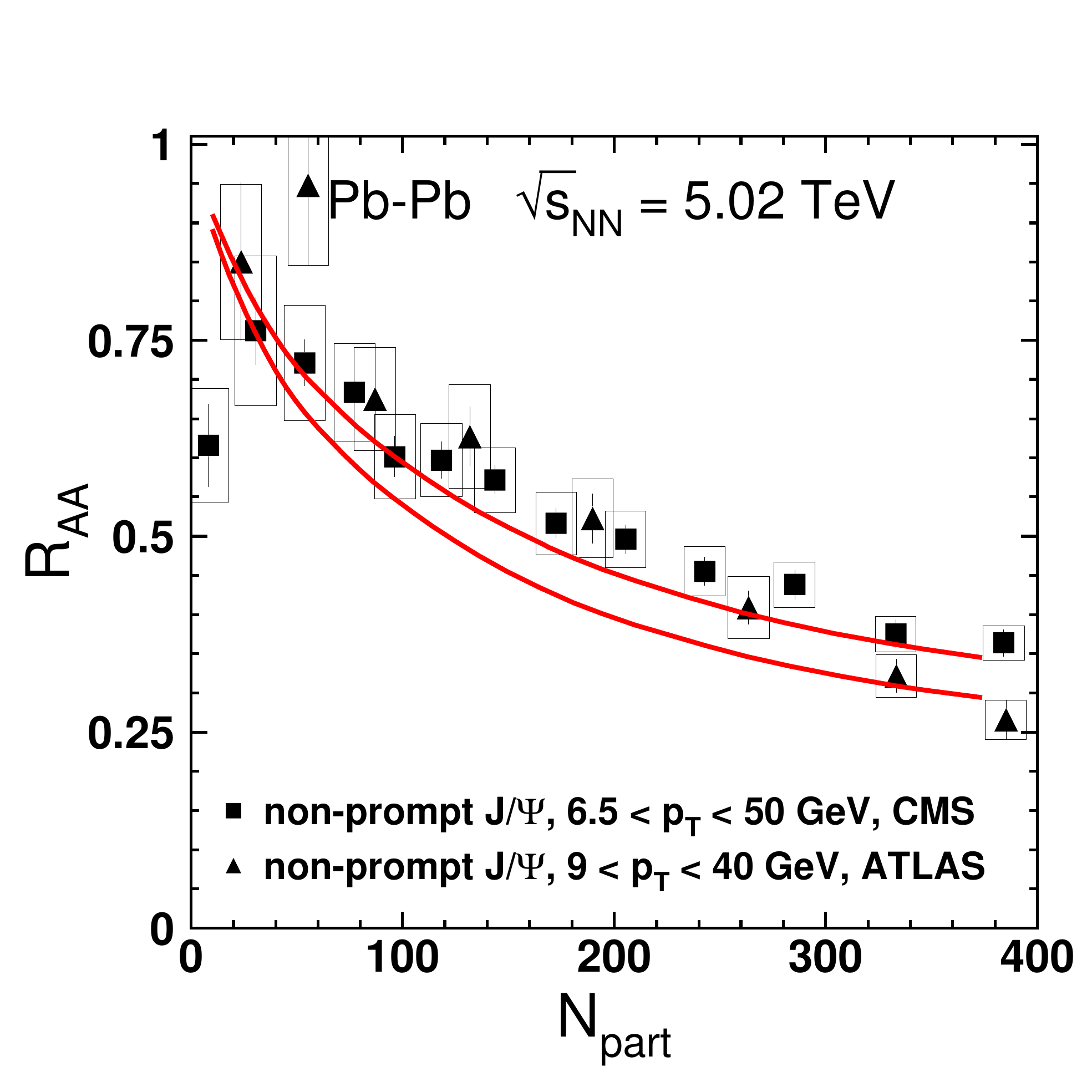}\vspace{-0.1cm}
\caption{Comparison with ATLAS \cite{b-atlas} and CMS \cite{cms-B-pp} data for indirect $J/\psi$ production            for minimum bias (or close to it) events as function of $p_T$ and vs centrality 
         at $\sqrt{s}=5.02\,\TeV$. The shown band corresponds to the uncertainty interval $\hat q_0= 0.2-0.25\GeV^2/\fm$. }
\label{fig:b-atlas-cms}
\end{figure}

 Comparison of calculated nuclear ratio $R_{AA}$, Eq.~(\ref{850})
 for $B$-meson production 
 with data from ATLAS \cite{b-atlas} and CMS \cite{cms-B-pp} experiments on indirect production of $J/\psi$
 from $B$ decays at $\sqrt{s}=5.02\,\TeV$ is depicted in Fig.~\ref{fig:b-atlas-cms} for minimum bias events
 vs $p_T$ and vs centrality.
 The value of the parameter $\hat q_0$ in (\ref{900}) required to describe data, is found amazingly small, $\hat q_0= 0.2-0.25\GeV^2/\fm$, nearly order of magnitude smaller than for light quarks (see next section).

\begin{figure}[h]
\centering 
\includegraphics[width=8.0cm,clip]{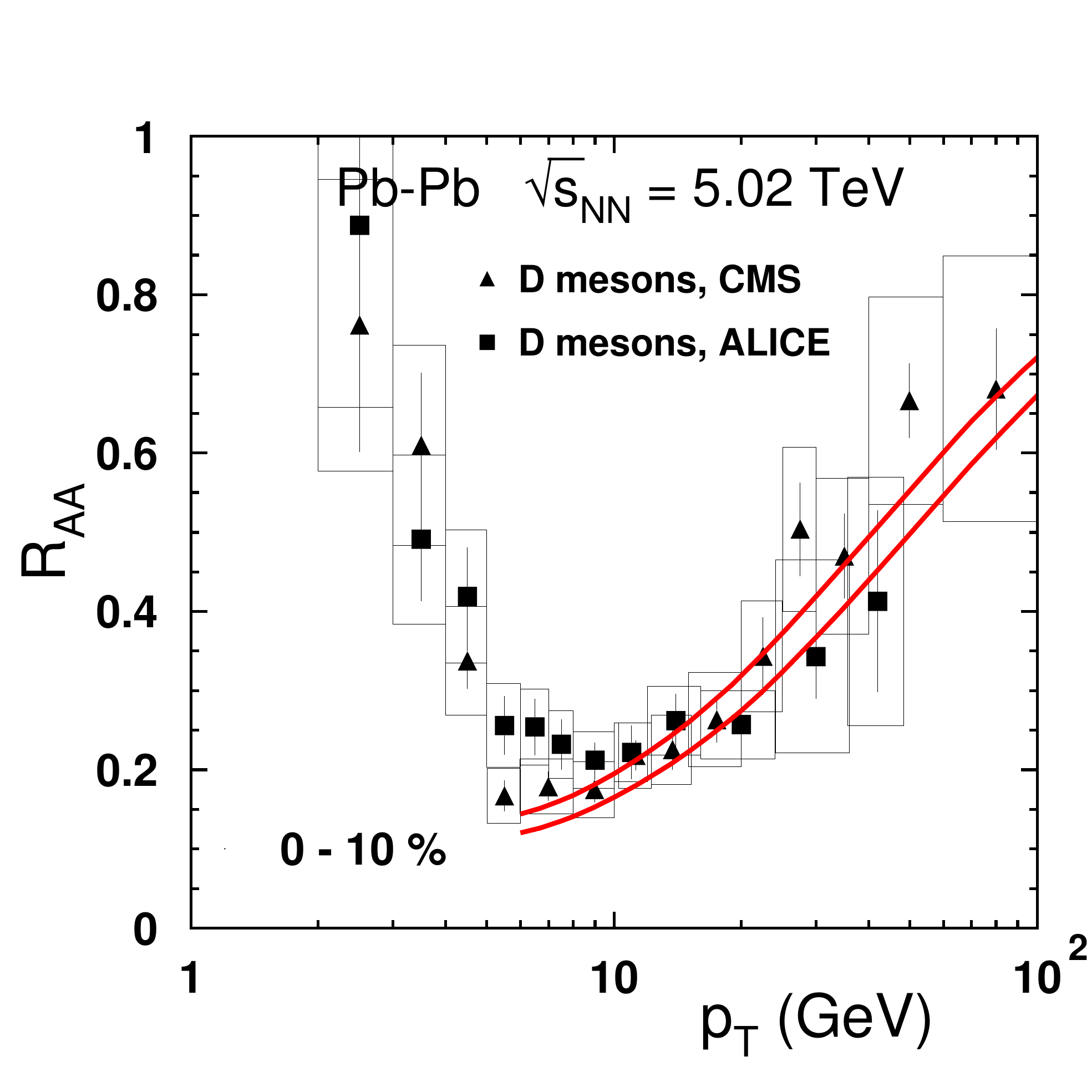}\vspace{-0.2cm}
\includegraphics[width=8.0cm,clip]{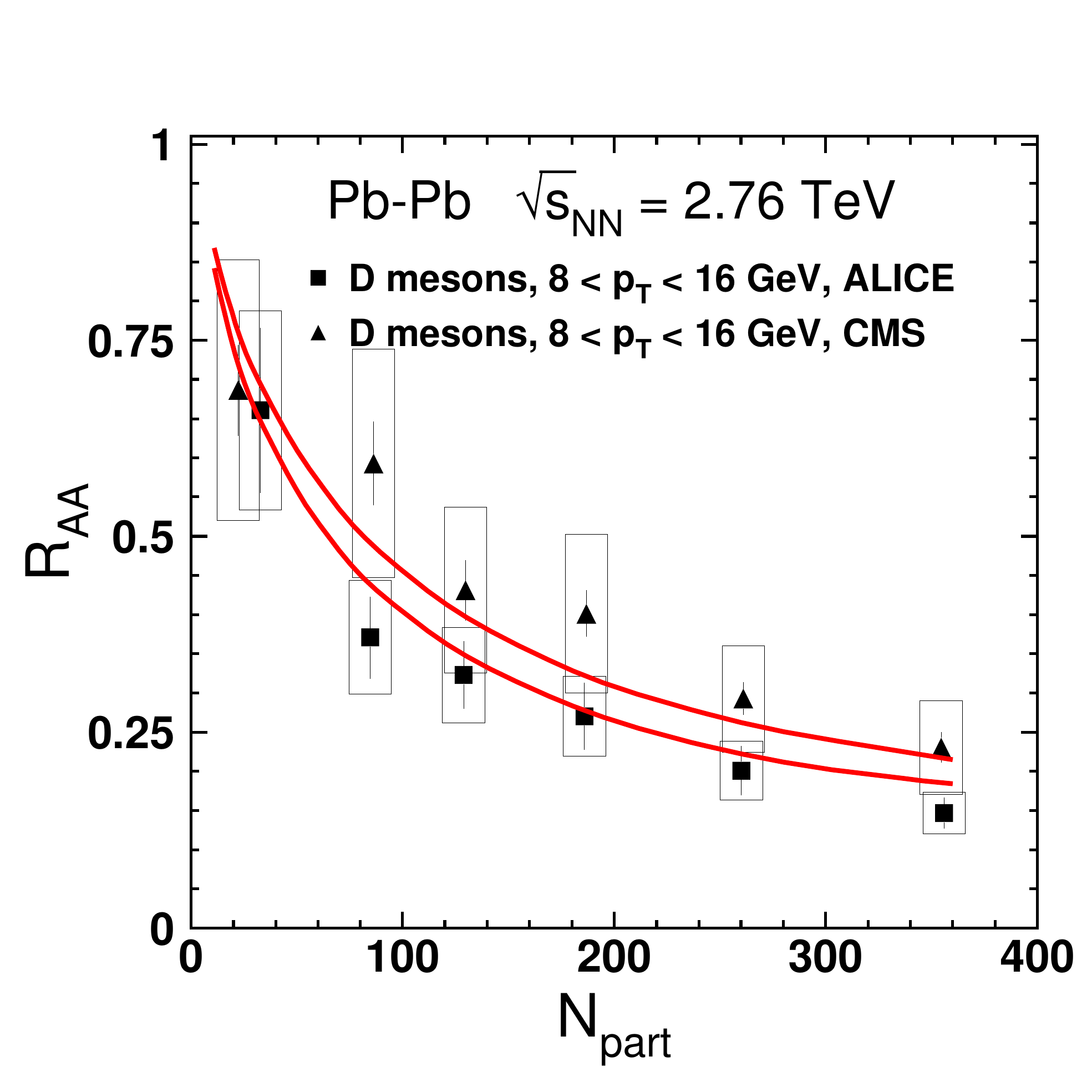}\vspace{-0.1cm}
\caption{Comparison with ALICE \cite{D-alice} and CMS \cite{cms-D-pp} data for indirect $D$-meson production 
         production at $\sqrt{s}=5.02\,\TeV$. The shown band corresponds to the uncertainty interval $\hat q_0= 0.45-0.55\GeV^2/\fm$.}
\label{fig:c-alice-cms}
\end{figure}

 The approach developed here can also be applied to production of $D$-mesons. The results are compared with data in Fig.~\ref{fig:c-alice-cms} vs $p_T$ and centrality.

 Notice that $c$-quarks radiate in vacuum much more energy 
 than $b$-quarks, while the effects of absorption 
 of $c\bar q$ and $b\bar q$ dipoles in the medium are similar. 
 Therefore, $D$-mesons are suppressed in $AA$ collisions 
 more than $B$-mesons.

 Another remarkable distinction is a flat $p_T$-dependence of $R_{AA}(p_T)$ for $B$-mesons 
 compared with the steep rise of $D$-meson rate observed in our calculations, confirmed with data. 
 This is a result of the interplay of the effects of color transparency (CT) and formation length.

 The observed steep rise with $p_T$ of $R_{AA}(p_T)$ for light hadrons and $J/\psi$ 
 is well explained by the CT effect \cite{knps}. Indeed, the initially produced small-size dipole is then expanding making the medium more opaque. At small $p_T$, which is the momentum of the dipole in the medium rest frame, the time of expansion, i.e. formation time is short, causing strong suppression of production rate. However, at large $p_T$ Lorentz time dilation "freezes" the small initial dipole size 
 for a longer time interval, making the medium more transparent due to CT. This is 
 why $R_{AA}(p_T)$ steeply rises.

 However, a heavy-light $Q\bar q$ dipole expands quickly to large hadronic size, 
 on a short path length given by (\ref{Lf}). Then, CT has no effect on the 
 survival of the dipole, this is why $R_{AA}(p_T)$ is levelling off.
 Nevertheless, the formation length Eq.~(\ref{Lf}) rises with $p_T$ and
 at higher $pT$ CT is again at work and $R_{AA}$ starts rising.

 If the typical path length in the medium is $L_{path}\sim5\fm$, 
 the transverse momenta required to reach the comparable value 
 $L_f\sim L_{path}$ are $p_T\gsim 40\GeV$ for $b$-quarks and $p_T\gsim 10\GeV$ for $c$-quarks. 
 Thus, onset of the CT regime for $B$-mesons is delayed compared with $D$-mesons, 
 as we see in upper plots of Figs.~\ref{fig:b-atlas-cms} and \ref{fig:c-alice-cms}.

\subsection{Transport coefficient for heavy flavors}\label{q-hat}
 The values of $\hat q_0$ required to adjust the results of calculations to data depicted in  
 Figs.~\ref{fig:b-atlas-cms}, \ref{fig:c-alice-cms} are 
 $\hat q^b_0= 0.20-0.25\GeV^2/\fm$ for $B$-meson and 
 $\hat q^c_0= 0.45-0.55\GeV^2/\fm$ for $D$-meson production. 
 This is significantly less than was found from data on high-$p_T$ production of light hadrons, 
 $\hat q_0= 2\GeV^2/\fm$ and $\hat q^q_0= 1.6\GeV^2/\fm$ at $\sqrt{s}=2.76\TeV$ and $200\GeV$
 respectively \cite{knps}. We see that the transport coefficient, 
 measured by heavy quarks, is remarkably low. 

 Reduction of $\hat q$ probed by charm has already revealed earlier.
 The analyses of $J/\psi$ suppression performed in \cite{kps-psi,k-psi} found
 $\hat q^c_0= 0.3-0.5\GeV^2/\fm$ in gold-gold collisions at $\sqrt{s}=200\GeV$, 
 which is quite consistent with present results for $\hat q_0$ at LHC. 
 Such an agreement looks nontrivial, since the mechanisms of suppression of 
 $J/\psi$ and $D$-mesons have so little to do in common. 

 Apparently the transport coefficient is not an universal characteristics of the hot medium 
 created in heavy ion collisions, and for good reason.
The transport coefficient $\hat q$ is the rate of broadening of a parton propagating through the medium, therefore it depends on kinematics and vary for different species of probing partons. 
 In the dipole approach broadening is calculated as \cite{jkt},
\beq 
\hat q = \left.\nabla_r^2 \,\sigma_{Q\bar Q}(r)\right\vert_{r=0}
\rho,
\label{eq:q-hat}
 \eeq
 where $\rho$ is the density of the medium (QGP).

 The dipole cross section $\sigma_{Q\bar Q}$ is suppressed for heavy quarks
 for many reasons. First, the radiative part of the cross section, which provides its rise with energy, 
 is suppressed by the dead cone effect, as was discussed above. 
 The small-$k_T$ part of gluon spectrum, which is 
 divergent for light quarks and provide an essential contribution to rising energy dependence, 
 is suppressed for heavy flavors. 
 Even omitting the radiative part, in the Born approximation, the dipole cross section is  proportional to  $\alpha_s^2(4m_Q^2)$.
 The coupling is $\alpha_s=0.195$ for $b$ and $\alpha_s=0.27$ for $c$-quarks \cite{brodsky}. 
 Thus we get $\sigma_{b\bar b}$ twice suppressed compared with $\sigma_{c\bar c}$. 
 This difference well explains the hierarchy of $\hat q_0$ values for heavy quarks found in our analysis. 

 Comparison with light quarks is more involved, because the QCD coupling is unknown at low scale. 
 In the Gribov’s theory of confinement \cite{Gribov:1998kb,Gribov:1999ui} the  distance  
 $0.3 \fm$ should correspond to the critical regime related to breaking of chiral symmetry. 
 Namely, at smaller distances, a perturbative quark-gluon basis is appropriate, 
 while at larger separations quasi-Goldstone pions emerge. 
 The corresponding critical value of the QCD running coupling was evaluated at $\alpha_s=0.43$ 
 \cite{Gribov:1998kb,Gribov:1999ui}.

 The standard phenomenological way to extend $\alpha_s(Q^2)$ down to small values $Q^2\to0$ 
 is to make a shift in the argument, $Q^2\Rightarrow Q^2+Q_0^2$ . 
 The value $Q_0^2=0.25\GeV^2$ was estimated in \cite{Dokshitzer:1995qm}. 
 This leads to the soft limit of $\alpha_s$ close to the above critical value.

 With this value of $\alpha_s$  the magnitudes of $\hat q_0$ for $b$ and $c$ quarks are suppressed 
 in comparison with light quarks by factors 5 and 2.5 respectively.
 This agrees pretty well with the extracted values of $\hat q_0$.
 Notice that the radiational part of the light quark dipole cross section, $\sigma_{q\bar q}(r)$ is not 
 negligible. Its addition should further improve
 the relations between transport coefficients probed by light and heavy quarks.

\section{Summary}
The present study of heavy quarks production in heavy ion collisions led to significant observations and  results.
\begin{itemize}

\item
 Gluon radiation by a heavy quark traced as function of path length, 
 ceases shortly on a distance of the order of $1\fm$ (depending on energy).

\item
 The full amount of radiated energy by a heavy quark is a small fraction of the quark energy.
 Therefore $B$-meson has to carry the main fraction of the $B$-jet momentum.
 Indeed the fragmentation function $D_{b/B}(z)$ measured in $e^+e^-$ annihilation 
 peaks at large $z\to1$. Similar behavior was observed for charm.
 
\item
 After color neutralization a colorless dipole $b\bar q$ does not radiate anymore.
 The production length $L_p$ distribution is directly related to the experimentally measured fragmentation function $D_{b/B}(z)$, so is unambiguously determined. The mean production length turned out to be extremely short, fractions of $\fm$. Therefore $B$ meson ($b\bar q$ dipole) is produced perturbatively.

\item
 Important consequence of shortness of $L_p$ is factorization of the production
 process to the independent stages of perturbative creation of a $b\bar q$ dipole and its FSI in the hot matter.
 
\item
 The only unknown of this analysis, and its goal, is the transport coefficient $\hat q$.
 Its maximal value was found $\hat q_0=0.2-0.25\GeV^2/\fm$ for $B$-mesons, and
 $\hat q_0=0.45-0.55\GeV^2/\fm$ for $D$-mesons.

\item
 These values are several times less than what was found in analysis of high-$p_T$ light hadron. 
 We explain this by running QCD coupling to a higher scale, and by
 dead-cone suppression of the radiative part of $\hat q$.
 
\end{itemize}

\section{Acknowledgement}
 This work was supported in part by grants CONICYT - Chile FONDECYT 1170319 and 1180232, by USM-TH-342 grant, and 
 by CONICYT - Chile PIA/BASAL FB0821.
 J.N. work was partially supported by Grants LTC17038 and LTT18002 
 of the Ministry of Education, Youth and Sports of the Czech Republic, 
 by the project of the European Regional Development Fund 
 CZ02.1.01/0.0/0.0/16\_019/0000778, 
 and by the Slovak Funding Agency, Grant 2/0007/18.


\begin{thebibliography}{}

\bibitem{kt-hf}
B.~Z.~Kopeliovich and A.~V.~Tarasov;
   ``Gluon shadowing and heavy flavor production off nuclei,''
   Nucl.\ Phys.\ A\textbf{710}, 180 (2002).

5~\bibitem{broad}
B.~Z.~Kopeliovich, I.~K.~Potashnikova and I.~Schmidt;
   ``Measuring the saturation scale in nuclei,''
   Phys.\ Rev.\ C\textbf{81}, 035204 (2010).

\bibitem{Lp}
B.~Z.~Kopeliovich, J.~Nemchik, I.~K.~Potashnikova and I.~Schmidt,
``Distinctive features of hadronizing heavy quarks,''
  arXiv:1909.08831 [hep-ph].

\bibitem{dk} 
Y.L.~Dokshitzer and D.E.~Kharzeev;
   ``Heavy quark colorimetry of QCD matter,''
   Phys. Lett. B\textbf{519}, 199 (2001).

\bibitem{b-atlas} 
M.~Aaboud et al. [ATLAS Collaboration];
   ``Prompt and non-prompt $J/\Psi$ and $\Psi'(2S)$ suppression
     at high transverse momentum in $5.02\TeV$ $Pb$+$Pb$
     collisions with the ATLAS experiment,''
   Eur. Phys. J. C\textbf{78}, 762 (2018).

\bibitem{D-alice} 
S.~Acharya {\it et al.} [ALICE Collaboration];
   ``Measurement of $D^0$, $D^+$,
     $D^{*+}$ and $D_{s}^+$ production in $Pb$-$Pb$ collisions at
       $\sqrt{s_{NN}}= 5.02\TeV$,''
    JHEP \textbf{1810}, 174 (2018).

\bibitem{bdmps} 
 R.~Baier, Y.~L.~Dokshitzer, A.~H.~Mueller, S.~Peigne and D.~Schiff,
 ``Radiative energy loss and p(T) broadening of high-energy partons in nuclei,''
  Nucl.\ Phys.\ B\textbf{484}, 265 (1997)

\bibitem{jet-lag} 
B.~Z.~Kopeliovich, H.-J.~Pirner, I.~K.~Potashnikova and I.~Schmidt;
   ``Jet lag effect and leading hadron production,''
   Phys.\ Lett.\ B\textbf{662}, 117 (2008).

\bibitem{within} 
B.~Z.~Kopeliovich, J.~Nemchik, E.~Predazzi and A.~Hayashigaki;
   ``Nuclear hadronization: Within or without?,''
   Nucl.\ Phys.\ A\textbf{740}, 211 (2004).

\bibitem{ct} 
B.~Z.~Kopeliovich, I.~K.~Potashnikova and I.~Schmidt;
   ``Color transparency and suppression of high-pT hadrons in nuclear collisions,''
   Phys.\ Rev.\ C\textbf{83}, 021901 (2011).

\bibitem{ct-eloss} 
B.~Z.~Kopeliovich, J.~Nemchik, I.~K.~Potashnikova and I.~Schmidt;
   ``Quenching of high-pT hadrons: Energy Loss vs Color Transparency,''
   Phys.\ Rev.\ C\textbf{86}, 054904 (2012)

\bibitem{troyan} 
Y.~L.~Dokshitzer, V.~A.~Khoze and S.~I.~Troian;
   ``On specific QCD properties of heavy quark fragmentation ('dead cone'),''
   J.\ Phys.\ G\textbf{17}, 1602 (1991).

\bibitem{knp}
B.Z.~Kopeliovich, J.~Nemchik and E.~Predazzi; in   
  \textit{Future Physics at HERA}, Proceedings of the Workshop 1995/96,
  edited by G. Ingelman, A. De Roeck and R. Klanner, DESY,
  1995/1996, vol.2, p. 1038; arXiv:\textbf{nucl-th/9607036}.

\bibitem{similar}
B.Z.~Kopeliovich, I.K.~Potashnikova and I.~Schmidt;
   ``Why heavy and light quarks radiate energy with similar rates,''
   Phys. Rev. C\textbf{82}, 037901 (2010).

\bibitem{charm} 
T.~Kneesch, B.A.~Kniehl, G.~Kramer and I.~Schienbein;
   ``Charmed-meson fragmentation functions with finite-mass corrections,''
   Nucl. Phys. B\textbf{799}, 34 (2008).

\bibitem{bottom}  
B.A.~Kniehl, G.~Kramer, I.~Schienbein and H.~Spiesberger;
   ``Finite-mass effects on inclusive $B$ meson hadroproduction,''
   Phys. Rev. D\textbf{77}, 014011 (2008).

\bibitem{kkp} 
B.A.~Kniehl, G.~Kramer and B.~Potter;
   ``Testing the universality of fragmentation functions,''
   Nucl. Phys. B\textbf{597}, 337 (2001).

\bibitem{roman}
V.~P.~Goncalves, B.~Kopeliovich, J.~Nemchik, R.~Pasechnik and I.~Potashnikova;
   ``Heavy flavor production in high-energy $pp$ collisions: color dipole description,''
   Phys.\ Rev.\ D\textbf{96}, 014010 (2017).

\bibitem{string1}
H.~Ichie, H.~Suganuma and H.~Toki;
   ``QCD phase transition at finite temperature in the dual Ginzburg-Landau theory,''
   Phys. Rev. D\textbf{52}, 2944 (1995).
 
\bibitem{string2} 
H.~Toki, S.~Sasaki, H.~Ichie and H.~Suganuma;
   ``Chiral symmetry breaking in the dual Ginzburg-Landau theory,''
   Austral. J. Phys. \textbf{50}, 199 (1997).
 
\bibitem{string3} 
H.~Ichie, H.~Suganuma and H.~Toki;
   ``The System of multicolor flux tubes in the dual Ginzburg-Landau theory,''
   Phys. Rev. D\textbf{54}, 3382 (1996).

\bibitem{radius} 
C.W.~Hwang;
   ``Charge radii of light and heavy mesons,''
   Eur. Phys. J. C\textbf{23}, 585 (2002).

\bibitem{atlas-B-pp} 
M.~Aaboud {\it et al.} [ATLAS Collaboration];
   ``Measurement of quarkonium production in proton-proton
    collisions at 5.02 TeV with the ATLAS detector,''
   Eur.\ Phys.\ J.\ C\textbf{78}, no. 3, 171 (2018).

\bibitem{cms-B-pp} 
A.~M.~Sirunyan {\it et al.} [CMS Collaboration];
   ``Measurement of prompt and nonprompt charmonium suppression in $\text {PbPb}$ collisions at 5.02 
    $\,\text{Te}\text{V}$,''
   Eur.\ Phys.\ J.\ C\textbf{78}, no. 6, 509 (2018).

\bibitem{jet-b}
S.~Chatrchyan {\it et al.} [CMS Collaboration],
  ``Inclusive $b$-jet production in $pp$ collisions at $\sqrt{s}=7$ TeV,''
  JHEP {\bf 1204}, 084 (2012)
  
  \bibitem{jet-c1} 
  S.~Acharya {\it et al.} [ALICE Collaboration],
  ``Measurement of the production of charm jets tagged with D$^{0}$ mesons in pp collisions at $ \sqrt{\mathrm{s}}=7 $ TeV,''
  JHEP {\bf 1908}, 133 (2019)
  
  \bibitem{jet-c2} 
  A.~M.~Sirunyan {\it et al.} [CMS Collaboration],
  ``Measurements of the charm jet cross section and nuclear modification factor in pPb collisions at $\sqrt{{s}_{NN}}$ = 5.02 TeV,''
  Phys.\ Lett.\ B {\bf 772}, 306 (2017)
  
\bibitem{alice-D-pp}
S.~Acharya {\it et al.} [ALICE Collaboration];
   ``Measurement of D-meson production at mid-rapidity in pp collisions at ${\sqrt{s}=7}$  TeV,''
   Eur.\ Phys.\ J.\ C\textbf{77}, no. 8, 550 (2017).

\bibitem{cms-D-pp}
A.~M.~Sirunyan {\it et al.} [CMS Collaboration];
   ``Nuclear modification factor of D$^0$ mesons in PbPb collisions at  $\sqrt{s_\mathrm{NN}} = 5.02$ TeV,''
   Phys.\ Lett.\ B\textbf{782}, 474 (2018).

\bibitem{wang} 
X.F.~Chen, C.~Greiner, E.~Wang, X.N.~Wang and Z.~Xu;
   ``Bulk matter evolution and extraction of jet 
     transport parameter in heavy-ion collisions at RHIC,''
   Phys. Rev. C\textbf{81}, 064908 (2010).

\bibitem{baier} 
R.~Baier;
   ``Jet quenching,''
   Nucl. Phys. A\textbf{715}, 209 (2003).

\bibitem{knps}
B.~Z.~Kopeliovich, J.~Nemchik, I.~K.~Potashnikova and I.~Schmidt;
   ``Quenching of high-pT hadrons: Energy Loss vs Color Transparency,''
   Phys.\ Rev.\ C\textbf{86}, 054904 (2012).

\bibitem{kps-psi}
B.~Z.~Kopeliovich, I.~K.~Potashnikova and I.~Schmidt;
   ``$J/\Psi$ production in nuclear collisions: Theoretical approach to measuring the transport coefficient,''
   Phys.\ Rev.\ C\textbf{82}, 024901 (2010).

\bibitem{k-psi}
B.~Z.~Kopeliovich;
   ``Puzzles of $J/\Psi$ production off nuclei,''
   Nucl.\ Phys.\ A\textbf{854}, 187 (2011).

\bibitem{jkt} 
M.~B.~Johnson, B.~Z.~Kopeliovich and A.~V.~Tarasov;
   ``Broadening of transverse momentum of partons propagating through a medium,''
   Phys.\ Rev.\ C\textbf{63}, 035203 (2001).

\bibitem{brodsky}
A.~Deur, S.~J.~Brodsky and G.~F.~de Teramond;
   ``The QCD Running Coupling,''
   Prog.\ Part.\ Nucl.\ Phys.\\textbf{90}, 1 (2016).

\bibitem{Gribov:1998kb} 
V.N.~Gribov;
   ``QCD at large and short distances (annotated version),''
   Eur.\ Phys.\ J.\ C\textbf{10}, 71 (1999).

\bibitem{Gribov:1999ui} 
V.~N.~Gribov;
   ``The Theory of quark confinement,''
   Eur.\ Phys.\ J.\ C\textbf{10}, 91 (1999).

\bibitem{Dokshitzer:1995qm} 
Y.~L.~Dokshitzer, G.~Marchesini and B.~R.~Webber;
   ``Dispersive approach to power behaved contributions in QCD hard processes,''
   Nucl.\ Phys.\ B\textbf{469}, 93 (1996).


\end{thebibliography}
\end{document}